\documentclass{article}

\usepackage{arxiv}
\usepackage{amsmath}

\usepackage[utf8]{inputenc} 
\usepackage[T1]{fontenc}    
\usepackage{hyperref}       
\usepackage{url}            
\usepackage{booktabs}       
\usepackage{amsfonts}       
\usepackage{nicefrac}       
\usepackage{microtype}      
\usepackage{lipsum}		
\usepackage{graphicx}
\usepackage{natbib}
\usepackage{doi}
\usepackage{float}
\usepackage{graphicx} 
\usepackage{rotating}

\title{Risk-Aware Deep Reinforcement Learning for Dynamic Portfolio Optimization}

\author{
\href{https://orcid.org/0009-0004-3167-8647}{\includegraphics[scale=0.06]{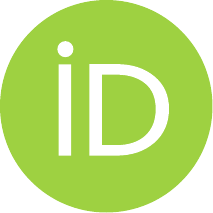}\hspace{1mm}Emmanuel Lwele}\thanks{Personal webpage: \url{https://emmanuellwele.com}} \\
Department of Engineering and Mathematics \\
Sheffield Hallam University \\
Sheffield S9 2AA \\
\texttt{e.lwele@shu.ac.uk}
	\And
	\href{https://orcid.org/0000-0000-0000-0000}{\includegraphics[scale=0.06]{orcid.pdf}\hspace{1mm}Sabuni Emmanuel}\thanks{Personal webpage: \url{https://www.sabuniemmanuelc.com/}} \\
Department of Financial Engineering \\
WorldQuant University \\
Washington, D.C. 2002 \\
\texttt{Sabuni0101@hotmail.com}
    \And
    \href{https://orcid.org/0000-0000-0000-0000}{\includegraphics[scale=0.06]{orcid.pdf}\hspace{1mm}Sitali Gabriel Sitali} \\
	Department of Financial Engineering\\
	WorldQuant University\\
	Washington, D.C. 20002 \\
	\texttt{sitaligabrielsitali@gmail.com} \\
}

\renewcommand{\shorttitle}{\textit{arXiv} Template}

\hypersetup{
pdftitle={Risk-Aware Deep Reinforcement Learning for Dynamic Portfolio Optimization},
pdfsubject={q-fin, q-fin.CP, q-fin.RM, cs.AI, q-fin.MF},
pdfauthor={Emmanuel Lwele, Sabuni Emmanuel},
pdfkeywords={Deep Reinforcement Learning, Dynamic Portfolio Allocation, Risk-Aware Optimization, Sharpe Ratio, Proximal Policy Optimization, Hidden Markov Models, Backtesting, Financial Time Series},
}

\begin{document}
\maketitle

\begin{abstract}
	This paper presents a deep reinforcement learning (DRL) framework for dynamic portfolio optimization in the presence of market uncertainty and risk. Traditional portfolio allocation methods often fall short in adapting to evolving market conditions while managing risk effectively. We introduce a risk-aware DRL model that integrates advanced reinforcement learning algorithms, a novel reward function based on the Sharpe ratio, and direct risk control strategies including maximum drawdown and portfolio volatility constraints. The model is trained and validated on historical financial data using Proximal Policy Optimization (PPO), and compared against mean-variance and equal-weighted benchmarks. Our empirical results show the DRL-based approach offers adaptive asset allocation and competitive performance under realistic constraints. This research contributes to the development of intelligent and risk-sensitive financial decision-making tools.
\end{abstract}


\section{Introduction}
Deep reinforcement learning (DRL) has emerged as a promising approach in quantitative finance, offering dynamic decision-making under uncertainty. Portfolio optimization, a core task in financial management, involves allocating assets to maximize returns while minimizing risk. Traditional methods based on modern portfolio theory (MPT) often fail to account for market dynamics and nonlinearities. The finance sector has been at the forefront of implementing cutting-edge technologies to enhance decision-making processes, and DRL has gained significant attention for its ability to adapt to complex, non-linear environments, particularly in dynamic portfolio allocation \cite{jankova2021bibliometric} \cite{zhong2019predicting}. This study proposes a DRL framework that incorporates risk-aware strategies to improve portfolio robustness and adaptability. It also examines the research landscape on integrating DRL techniques in finance, focusing on how risk awareness is incorporated into the optimization process.

\section{Related Work and Project Objectives}
\label{sec:relatedWorks}
This project aims to develop a specialized Deep Reinforcement Learning (DRL) model for dynamic portfolio allocation within the finance sector. Traditional methods often struggle to adapt to changing market conditions and effectively manage risk. By incorporating risk awareness into the optimization process, this project seeks to overcome these limitations.

Historical data was utilised for model training and backtesting, with monthly portfolio rebalancing. Key objectives included designing a unique reward function, exploring advanced DRL methodologies, integrating risk management techniques, conducting comprehensive backtesting, performing sensitivity analysis, and deriving actionable insights for investors and portfolio managers. This research intends to make distinctive contributions by exploring innovative strategies and novel methodological combinations, thereby advancing the application of DRL in risk-aware portfolio optimization.

\subsection{Goals and Objectives}
\begin{enumerate}
    \item Develop a DRL system tailored for dynamic portfolio allocation, enabling adaptive asset allocations over time.
    \item Design and implement a novel reward function incorporating a risk-adjusted performance metric, such as the Sharpe ratio, to balance risk management with return maximisation.
    \item Integrate risk management techniques directly into the DRL framework, including constraints on maximum drawdown, portfolio volatility, and other relevant risk measures, to ensure robustness and resilience.
\end{enumerate}

Recent literature explores the intersection of DRL and finance, particularly focusing on risk-sensitive approaches. Prior studies have leveraged DRL models to directly optimize financial performance metrics such as the Sharpe ratio and mean-variance objectives \cite{jankova2021bibliometric}. Notable advancements include distributional reinforcement learning and hybrid architectures that combine deep learning with classical statistical methods \cite{kilimci2020efficient}. However, many existing approaches lack the explicit integration of real-time risk controls and adaptive capabilities, underscoring the need for enhanced frameworks such as the one proposed in this project.

\section{Background}

Optimizing one's portfolio is a fundamental component of modern trading systems. The primary goal of portfolio optimization is to determine the most effective allocation of assets to maximize returns while maintaining a specified level of risk. This concept is encapsulated in Modern Portfolio Theory (MPT), developed by Harry Markowitz in his seminal work.

MPT emphasizes the importance of diversification, advocating for a balanced mix of assets to reduce overall portfolio risk. By spreading investments across a variety of instruments, diversification helps smooth out the equity curve, minimizing the impact of individual asset volatility. As a result, such a strategy tends to yield a higher return per unit of risk compared to investing in single assets. This foundational principle continues to influence contemporary approaches to asset management and portfolio design.

\section{Literature Review}

\subsection{Deep Reinforcement Learning in Finance}

Deep Reinforcement Learning (DRL) integrates reinforcement learning with deep learning, enabling high-performance solutions for complex sequential decision-making tasks. In the finance sector, DRL has been successfully applied to asset pricing, portfolio management, and algorithmic trading, offering the ability to adapt strategies to non-linear and stochastic market dynamics.

\subsection{Dynamic Portfolio Allocation}

Dynamic portfolio allocation involves the periodic rebalancing of investment portfolios to optimize returns while minimizing risk exposure. Traditional strategies often rely on static models or heuristic rules, which may be inadequate in responding to real-time market changes. DRL offers a powerful alternative by leveraging historical and current market data to learn optimal asset allocation policies dynamically.

\subsection{Incorporating Risk Awareness}

Effective risk management is essential in achieving financial goals while limiting exposure to adverse market events. Incorporating risk awareness directly into the DRL optimization process enhances portfolio resilience. This can be achieved by embedding risk measures, such as volatility or drawdown constraints, into the reward function.

Recent studies have proposed various risk-sensitive DRL frameworks. For example, \cite{zhong2019predicting} proposed a DRL model that directly optimizes the Sharpe ratio to balance risk and return. Similarly, \cite{8923223} introduced a Deep Deterministic Policy Gradient (DDPG)-based approach that incorporates risk sensitivity into the training objective. Hybrid models have also been explored; for instance, \cite{ghasemzadeha2020machine} combined DRL with mean-variance optimization.

Further advancements include Distributional Reinforcement Learning (DistRL), which focuses on optimizing the distribution of risk-adjusted returns. For example, \cite{10.5555/3600270.3602516} presented a distributional RL model for dynamic portfolio allocation, which enables more robust risk modeling by considering the full distribution of possible outcomes rather than just the expected return.

\subsection{Competitor Analysis}

Although there have been significant advancements in the deployment of DRL techniques for portfolio allocation, current models often overlook explicit risk management integration. Traditional optimization and heuristic methods struggle with adaptability in rapidly changing markets. These gaps present opportunities for developing more robust, risk-aware DRL frameworks.

\subsection{Business Case}

With the financial sector becoming increasingly data-driven and automated, advanced methods for adaptive portfolio optimization are essential. DRL provides a viable solution by enabling learning-based allocation strategies that adapt to evolving market conditions. By integrating risk awareness, the proposed model enhances decision-making capabilities and offers a competitive edge to financial institutions and investment managers.

\subsection{SWOT Analysis of Competitors}

Competitors' strengths lie in their optimized methodologies, which are underpinned by robust theoretical foundations. However, key weaknesses include limited adaptability to rapidly changing market dynamics and the insufficient integration of risk management frameworks. These gaps offer significant opportunities for innovation in DRL-based portfolio allocation strategies. Threats to the adoption of such models include competition from established financial institutions and potential regulatory or legal challenges. Nevertheless, recent advancements in DRL systems—particularly those focused on risk-aware optimization—help mitigate these risks and position such frameworks as viable alternatives in modern financial decision-making.

\section{Research Design and Methodology}

In this section, we will introduce our framework and discuss how Sharpe ratio can be optimized through gradient ascent.

\subsection{Sharpe Ratio-Based Objective Function}

The Sharpe ratio, defined as the expected return over volatility (excluding the risk-free rate for simplicity), is used to quantify return per unit of risk for a portfolio:

\begin{equation}
L = \frac{E(R_p)}{\text{Std}(R_p)}
\end{equation}

where \( E(R_p) \) and \( \text{Std}(R_p) \) are the mean and standard deviation of portfolio returns, respectively.

For a trading period \( t = \{1, \ldots, T\} \), the time-dependent Sharpe ratio objective function \( L_t \) is defined as:

\begin{equation}
L_t = \frac{E(R_{p,t})}{\sqrt{E(R_{p,t}^2) - (E(R_{p,t}))^2}}
\end{equation}

The expected return over the period is computed as:

\begin{equation}
E(R_{p,t}) = \frac{1}{T} \sum_{t=1}^{T} R_{p,t}
\end{equation}

where \( R_{p,t} \) denotes the realized portfolio return at time \( t \), computed as:

\begin{equation}
R_{p,t} = \sum_{i=1}^{n} w_{i,t-1} \cdot r_{i,t}
\end{equation}

Here, \( r_{i,t} = \left( \frac{p_{i,t}}{p_{i,t-1}} - 1 \right) \) is the return of asset \( i \), and \( w_{i,t} \in [0,1] \) is the allocation weight of asset \( i \) at time \( t \), with the constraint \( \sum_{i=1}^{n} w_{i,t} = 1 \).

To model \( w_{i,t} \) for a long-only portfolio, a neural network \( f \) with parameters \( \theta \) is used, which takes current market features \( x_t \) as input:

\begin{equation}
w_{i,t} = f(\theta \mid x_t)
\end{equation}

Where \( x_t \) represents the current market information, and the classical forecasting step was bypassed by directly linking the inputs to position weights in order to maximize the Sharpe ratio over the trading period \( T \), namely \( L_T \). 

However, since a long-only portfolio was considered, the weights were required to satisfy the constraints of being positive and summing to one. To enforce these constraints, the softmax function was applied to the raw outputs of the neural network:

\begin{equation}
w_{i,t} = \frac{\exp(\bar{w}_{i,t})}{\sum_{j=1}^{n} \exp(\bar{w}_{j,t})}
\end{equation}

where \( \bar{w}_{i,t} \) is the unnormalized output (logit) of the neural network for asset \( i \) at time \( t \).

This formulation ensures all weights \( w_{i,t} \in (0,1) \) and \( \sum_{i=1}^{n} w_{i,t} = 1 \), thus conforming to portfolio constraints.

Unconstrained optimization techniques can be applied to maximize the Sharpe ratio-based objective function described earlier. In particular, the \textbf{gradient ascent} was employed to optimize the model parameters.

An analytical derivation of the gradient of \( L_T \) with respect to the neural network parameters \( \theta \) can be found in \cite{molina2016stock, moody1998performance}. Once the gradient \( \nabla_\theta L_T \) is computed, the parameters are updated using the following gradient ascent rule:

\begin{equation}
\theta_{\text{new}} = \theta_{\text{old}} + \alpha \cdot \frac{\partial L_T}{\partial \theta}
\end{equation}

where \( \alpha \) is the learning rate. This update step is repeated iteratively over training epochs until convergence is achieved or the Sharpe ratio on the validation data is optimized.

\subsection{Efficient Frontier}

An efficient frontier is a graph where the y-axis represents the expected return and the x-axis represents the portfolio's risk, typically measured by the standard deviation of portfolio returns. This tool visualizes efficient, global minimum variance, and inefficient portfolios, as well as the overall risk-return trade-off of investment strategies.

The efficient frontier was employed as a decision-support tool to facilitate more informed investment choices, as it enabled the identification of portfolios that offered higher expected returns with comparatively lower levels of risk.

The expected return of a two-asset portfolio, a foundational concept in constructing the efficient frontier, is calculated as:

\begin{equation}
E(R_p) = w_1 E(R_1) + w_2 E(R_2)
\end{equation}

where \( w_1 \) and \( w_2 \) are the respective weights of asset 1 and asset 2 in the portfolio.

Graphically, the efficient frontier forms the upper edge of the feasible set of portfolios in risk-return space, showing optimal portfolios that offer the maximum expected return for a given level of risk.

\begin{figure}[H]
    \centering
    \includegraphics[width=0.7\textwidth]{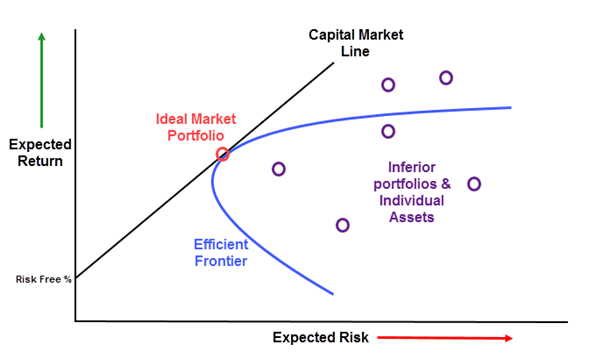}
    \caption{Efficient Frontier illustrating the trade-off between expected return and risk. Portfolios on the frontier dominate inferior portfolios in terms of return per unit risk. The Capital Market Line (CML) shows the optimal portfolio combinations including a risk-free asset.}
    \label{fig:efficient_frontier}
\end{figure}

\subsection{Hidden Markov Models}

The Hidden Markov Model (HMM) approach is applied to the stock prediction problem due to its strong ability to model temporal dependencies. HMMs have shown success in analyzing and forecasting time-dependent phenomena, particularly in the context of time series data \cite{basak2019predicting,reddy2018stock}.

Given the inherent sequential and stochastic nature of stock market data, HMMs offer a suitable framework for capturing the hidden states of market conditions that influence observable variables like asset prices. By modeling these hidden states, HMMs can be used to predict future movements in the market, aiding in more informed decision-making for portfolio allocation.

\subsection{Deep Reinforcement Learning Models (DRL)}

The Deep Reinforcement Learning (DRL) model performed dynamic asset allocation using the Proximal Policy Optimization (PPO) algorithm. PPO is a policy gradient method known for its stability and sample efficiency, making it well-suited for financial applications where data is sequential and policy improvements must be carefully constrained to avoid overfitting and poor generalization.

The DRL agent learns a policy that maps observed market states to portfolio weight allocations, with the goal of maximizing a custom-defined reward function based on risk-adjusted returns. PPO ensures that each update stays within a trust region, reducing the chances of large policy shifts that could destabilize learning in volatile financial environments.

\subsection{Deep Neural Networks Framework}

A Deep Neural Network (DNN) is proposed as the agent in our DRL framework. It is responsible for learning an optimal policy that maps observed market states to actions, which in this context are asset allocation decisions. The network is trained to maximize a reward function that reflects risk-adjusted portfolio performance.

The DRL process is modeled as a standard reinforcement learning loop, where:
\begin{itemize}
    \item The \textbf{state} represents the current market conditions, including features such as historical asset prices, technical indicators, and volatility estimates.
    \item The \textbf{action} corresponds to a portfolio weight vector that determines the allocation across different assets.
    \item The \textbf{reward} is derived from the resulting portfolio performance, adjusted for risk and transaction costs.
    \item The \textbf{policy} is a function approximator, implemented as a deep neural network, which aims to choose actions that maximize the expected cumulative reward.
\end{itemize}

This neural network-based policy is optimized using PPO (Proximal Policy Optimization), allowing the agent to gradually improve its decision-making over time by interacting with the market environment simulated through historical data.

\begin{figure}[htbp]
    \centering
    \includegraphics[width=0.7\textwidth]{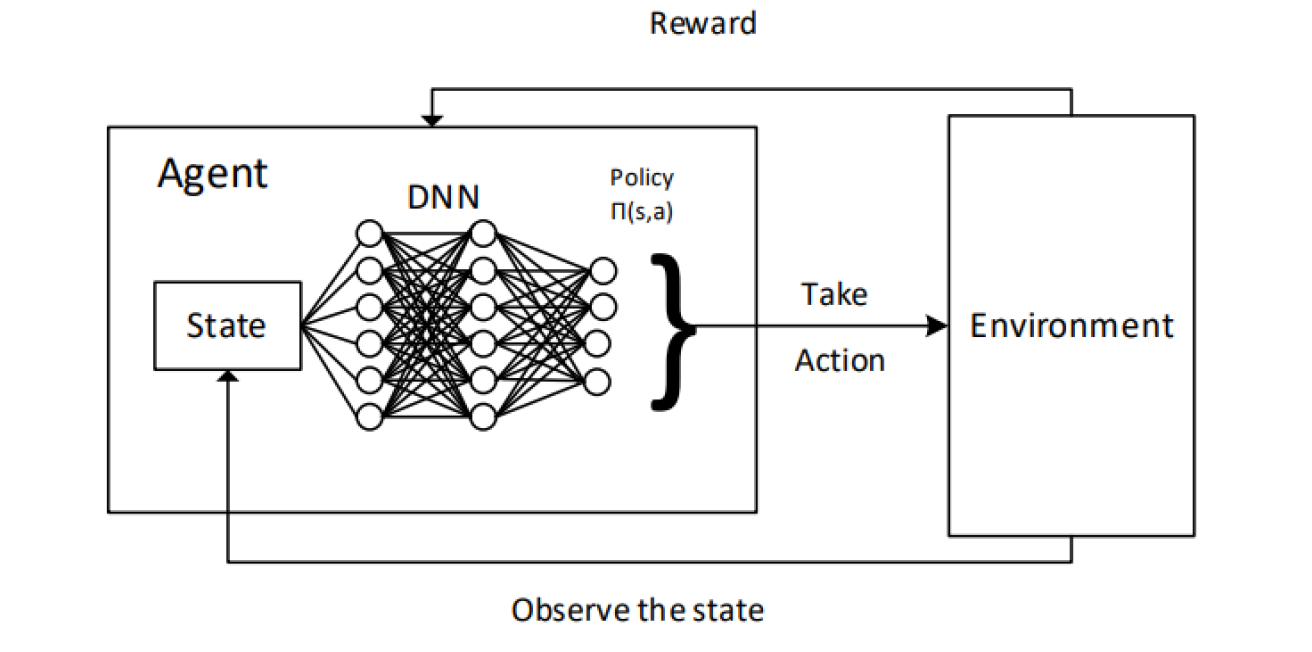}
    \caption{Basic representation of a Deep Reinforcement Learning (DRL) framework. The agent observes the state from the environment and uses a deep neural network (DNN) to compute a policy \( \pi(s, a) \). Based on the policy, the agent takes an action, receives a reward, and updates its policy to improve future decisions.}
    \label{fig:drl_framework}
\end{figure}

\begin{figure}[htbp]
    \centering
    \includegraphics[width=0.7\textwidth]{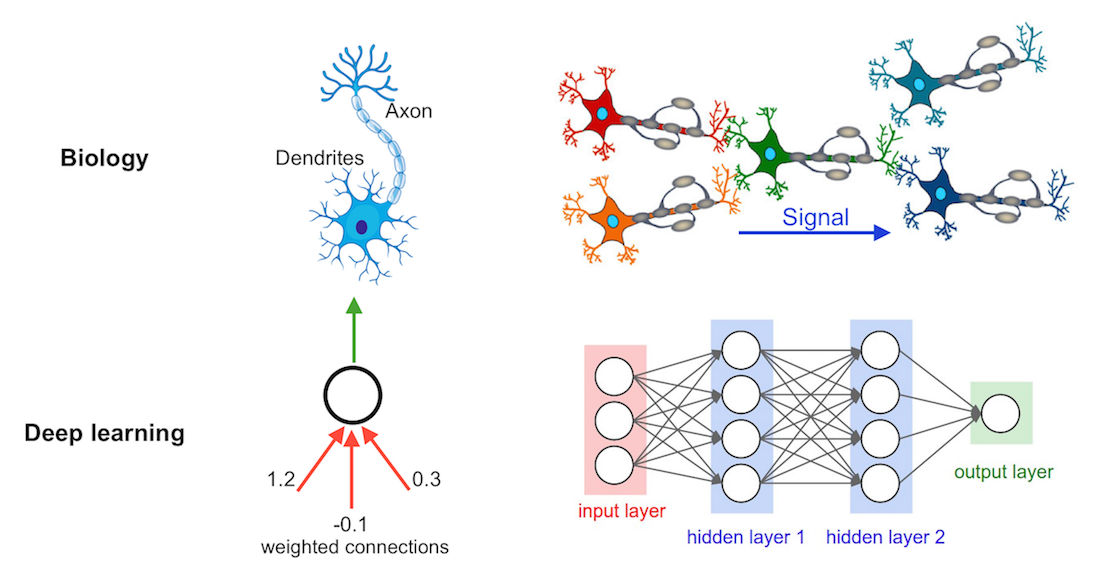}
    \caption{Comparison between biological neural networks and deep learning architectures. The biological neuron (top left) is the inspiration for artificial neurons in deep learning (bottom left). Signals in biological networks are passed between neurons (top right), similar to how inputs are processed through layers of a deep neural network (bottom right).}
    \label{fig:dnn_biology_comparison}
\end{figure}

\section{Data Preparation and Model Configuration}

\subsection{Data Preparation}

Historical financial data was collected, including stock prices, economic indicators, and other relevant financial features. The data underwent preprocessing steps such as handling missing values, normalization or standardization of input features, and splitting into training, validation, and test datasets.

\subsection{Model Architecture}

The Reinforcement Learning (RL) configuration of our DRL model is defined as follows:

\begin{enumerate}
    \item \textbf{State (\( S \))}: The state includes features such as the high, low, and close prices, the covariance matrix of closing prices, and selected market indicators used as model inputs.
    
    \item \textbf{Action (\( A \))}: The action at time \( t \), denoted as \( A_t \), is the portfolio weight vector, derived from the weights at the previous time step \( A_{t-1} = W_{t-1} \). The agent decides on \( W_t \), considering transaction costs, to maximize cumulative portfolio value.
    
    \item \textbf{Reward Function}: The reward at time \( t \) is based on the log of the total portfolio value, penalized by transaction costs. It is defined as:
    \begin{equation}
    R(S_{t-1}, A_{t-1}) = \ln \left( A_{t-1} \cdot Y_{t-1} - \mu \sum_{i=1}^{n} \left| A_{i,t-1} - W_{i,t-1} \right| \right)
    \end{equation}
    where \( Y_{t-1} \) represents the relative price change vector, and \( \mu \) is the transaction cost coefficient.
    
    \item \textbf{Policy (\( \pi(s, a) \))}: The policy determines the optimal action at each state and is implemented using a deep neural network trained with the PPO algorithm.
\end{enumerate}

\subsection{Training and Loss Function}

The Deep Neural Network was trained on the training dataset using Proximal Policy Optimization (PPO). The loss function captures the portfolio allocation performance, incorporating either standard mean squared error (MSE) or a custom loss reflecting risk-adjusted returns.

The MSE defined as:

\begin{equation}
\text{MSE} = \frac{1}{n} \sum_{i=1}^{n} (y_i - \hat{y}_i)^2
\end{equation}

Model parameters were optimized using gradient-based optimizers such as Stochastic Gradient Descent (SGD), Adam, and RMSprop.

\subsection{Validation and Hyperparameter Tuning}

Model validation was performed on a separate validation set, using metrics like Sharpe ratio, cumulative returns, and portfolio volatility. Hyperparameters (e.g., learning rate, batch size, number of epochs) were tuned using grid search and random search techniques.

\subsection{Evaluation}

The trained model was evaluated on the test dataset to assess generalization performance and was benchmarked against traditional portfolio strategies, including mean-variance optimization and equal-weighted portfolios.

\subsection{Deployment}

Following satisfactory performance, the trained model was deployed for real-time portfolio allocation. The system continues to monitor its performance and is periodically retrained to adapt to evolving market conditions.

\section{Overfitting and Non-Stationarity}

Given that historical financial data is inherently time series in nature, two key challenges arise: overfitting and non-stationarity. To ensure robust model performance, the following mitigation strategies were applied:

\subsection{Overfitting Mitigation}

\begin{itemize}
    \item \textbf{Cross-validation}: The dataset was partitioned into training and validation sets, and techniques such as k-fold cross-validation were employed to assess the model’s ability to generalize to unseen data.
    \item \textbf{Regularization}: We applied regularization techniques such as L1 (Lasso) and L2 (Ridge) to penalize large weights in the neural network and reduce the risk of overfitting.
\end{itemize}

\subsection{Handling Non-Stationarity}

\begin{itemize}
    \item \textbf{Seasonal Adjustment}: Seasonal decomposition methods were used to extract and remove seasonal patterns from the data.
    \item \textbf{Rolling Statistics}: Rolling averages and rolling standard deviations were calculated to smooth out time-based fluctuations and adapt to evolving trends in the data.
\end{itemize}

\section{Backtesting}

The performance of our DRL model against traditional benchmark strategies was evaluated, including mean-variance optimized portfolios and equal-weighted portfolios. The DRL model was trained and executed using the Proximal Policy Optimization (PPO) algorithm. The comparison was carried out to assess the practical benefits of DRL-based asset allocation in real market environments.

\section{Empirical Results}

\subsection{Dataset}

For this experiment, 15 high-performing assets were selected from Yahoo Finance on the basis of their historical performance over a 10-year period spanning January 1, 2015, to January 1, 2024.Daily adjusted close prices were collected, and the expected returns for each asset were computed. A summary of the dataset was presented in Table~\ref{tab:asset-head-all}, which shows a snapshot of the selected assets.

\begin{table}[htbp]
\centering
\caption{Head of Adjusted Close Prices for All Assets}
\label{tab:asset-head-all}
\resizebox{\textwidth}{!}{%
\begin{tabular}{lrrrrrrrrrrrrrrr}
\toprule
\textbf{Date} & \textbf{AAPL} & \textbf{AMZN} & \textbf{GOOG} & \textbf{HD} & \textbf{INTC} & \textbf{JNJ} & \textbf{JPM} & \textbf{MA} & \textbf{META} & \textbf{NFLX} & \textbf{NVDA} & \textbf{PG} & \textbf{TSLA} & \textbf{UNH} & \textbf{V} \\
\midrule
2015-01-02 & 24.4022 & 15.4260 & 26.1687 & 83.4010 & 28.0734 & 80.5544 & 48.2683 & 80.7925 & 78.3669 & 49.8486 & 4.8326 & 69.0902 & 14.6207 & 87.5710 & 62.0182 \\
2015-01-05 & 23.7147 & 15.1095 & 25.6232 & 81.6513 & 27.7569 & 79.9918 & 46.7698 & 78.5200 & 77.1082 & 47.3114 & 4.7510 & 68.7618 & 14.0060 & 86.1286 & 60.6493 \\
2015-01-06 & 23.7170 & 14.7645 & 25.0293 & 81.4013 & 27.2396 & 79.5987 & 45.5571 & 78.3503 & 76.0693 & 46.5014 & 4.6069 & 68.4485 & 14.0853 & 85.9548 & 60.2584 \\
2015-01-07 & 24.0495 & 14.9210 & 24.9864 & 84.1912 & 27.8109 & 81.3560 & 45.6267 & 79.5690 & 76.0693 & 46.7429 & 4.5949 & 68.8076 & 14.0633 & 86.8324 & 61.0658 \\
2015-01-08 & 24.9736 & 15.0230 & 25.0652 & 86.0539 & 28.3282 & 81.9956 & 46.6463 & 80.8067 & 78.0971 & 47.7800 & 4.7678 & 69.5944 & 14.0413 & 90.9772 & 61.8848 \\
\bottomrule
\end{tabular}%
}
\end{table}

The dataset utilised  had 15 selected assets, namely: \texttt{AAPL}, \texttt{AMZN}, \texttt{GOOG}, \texttt{HD}, \texttt{INTC}, \texttt{JNJ}, \texttt{JPM}, \texttt{MA}, \texttt{META}, \texttt{NFLX}, \texttt{NVDA}, \texttt{PG}, \texttt{TSLA}, \texttt{UNH}, and \texttt{V}.

\begin{table}[htbp]
\centering
\caption{Covariance Matrix for All Assets}
\label{tab:cov-matrix}
\resizebox{\textwidth}{!}{%
\begin{tabular}{lrrrrrrrrrrrrrrr}
\toprule
Ticker & AAPL & AMZN & GOOG & HD & INTC & JNJ & JPM & MA & META & NFLX & NVDA & PG & TSLA & UNH & V \\
\midrule
AAPL & 0.084324 & 0.054972 & 0.051678 & 0.037654 & 0.051861 & 0.019914 & 0.035967 & 0.048541 & 0.060289 & 0.055630 & 0.079386 & 0.021634 & 0.070397 & 0.032933 & 0.043037 \\
AMZN & 0.054972 & 0.110670 & 0.062523 & 0.035056 & 0.047055 & 0.015116 & 0.027556 & 0.045135 & 0.074129 & 0.077962 & 0.082731 & 0.015604 & 0.074432 & 0.025565 & 0.039662 \\
GOOG & 0.051678 & 0.062523 & 0.081754 & 0.034988 & 0.048019 & 0.018200 & 0.034120 & 0.046950 & 0.070122 & 0.059988 & 0.075666 & 0.018567 & 0.060760 & 0.029775 & 0.042013 \\
HD   & 0.037654 & 0.035056 & 0.034988 & 0.061295 & 0.039991 & 0.018774 & 0.036153 & 0.037899 & 0.040449 & 0.036586 & 0.054154 & 0.020983 & 0.044634 & 0.030644 & 0.035197 \\
INTC & 0.051861 & 0.047055 & 0.048019 & 0.039991 & 0.111970 & 0.020810 & 0.041940 & 0.046560 & 0.055715 & 0.054340 & 0.082922 & 0.022478 & 0.061899 & 0.030882 & 0.041095 \\
JNJ  & 0.019914 & 0.015116 & 0.018200 & 0.018774 & 0.020810 & 0.033677 & 0.021164 & 0.021713 & 0.017090 & 0.014437 & 0.019798 & 0.019151 & 0.013739 & 0.023218 & 0.020406 \\
JPM  & 0.035967 & 0.027556 & 0.034120 & 0.036153 & 0.041940 & 0.021164 & 0.077114 & 0.045590 & 0.034884 & 0.029767 & 0.047775 & 0.019257 & 0.042006 & 0.033941 & 0.041597 \\
MA   & 0.048541 & 0.045135 & 0.046950 & 0.037899 & 0.046560 & 0.021713 & 0.045590 & 0.076418 & 0.052117 & 0.045131 & 0.068395 & 0.021789 & 0.055346 & 0.034337 & 0.062505 \\
META & 0.060289 & 0.074129 & 0.070122 & 0.040449 & 0.055715 & 0.017090 & 0.034884 & 0.052117 & 0.141398 & 0.078434 & 0.088671 & 0.017890 & 0.072368 & 0.027187 & 0.045246 \\
NFLX & 0.055630 & 0.077962 & 0.059988 & 0.036586 & 0.054340 & 0.014437 & 0.029767 & 0.045131 & 0.078434 & 0.201734 & 0.090949 & 0.015502 & 0.087610 & 0.030770 & 0.040811 \\
NVDA & 0.079386 & 0.082731 & 0.075666 & 0.054154 & 0.082922 & 0.019798 & 0.047775 & 0.068395 & 0.088671 & 0.090949 & 0.231897 & 0.022565 & 0.115494 & 0.039747 & 0.060201 \\
PG   & 0.021634 & 0.015604 & 0.018567 & 0.020983 & 0.022478 & 0.019151 & 0.019257 & 0.021789 & 0.017890 & 0.015502 & 0.022565 & 0.035424 & 0.014415 & 0.020543 & 0.020512 \\
TSLA & 0.070397 & 0.074432 & 0.060760 & 0.044634 & 0.061899 & 0.013739 & 0.042006 & 0.055346 & 0.072368 & 0.087610 & 0.115494 & 0.014415 & 0.318396 & 0.031284 & 0.050173 \\
UNH  & 0.032933 & 0.025565 & 0.029775 & 0.030644 & 0.030882 & 0.023218 & 0.033941 & 0.034337 & 0.027187 & 0.030770 & 0.039747 & 0.020543 & 0.031284 & 0.067703 & 0.032197 \\
V    & 0.043037 & 0.039662 & 0.042013 & 0.035197 & 0.041095 & 0.020406 & 0.041597 & 0.062505 & 0.045246 & 0.040811 & 0.060201 & 0.020512 & 0.050173 & 0.032197 & 0.063157 \\
\bottomrule
\end{tabular}%
}
\end{table}

\noindent The covariance was  calculated to establish the linear relationship amongst the assets and, using the efficient frontier, identified the most suitable assets. The portfolio risk (standard deviation) on the frontier is given by:

\begin{equation}
\sigma_p
= \sqrt{
\sum_{i=1}^{N} w_i^{2}\,\sigma_i^{2}
\;+\;
\sum_{i=1}^{N}\sum_{\substack{j=1\\ j\neq i}}^{N} w_i w_j \,\mathrm{Cov}(k_i,k_j)
}
\label{eq:portfolio-risk}
\end{equation}

\noindent where $w_i$ are asset weights, $\sigma_i^2$ are individual asset variances, and $\mathrm{Cov}(k_i,k_j)$ are pairwise return covariances.

\medskip
\noindent After applying this framework, the following assets were found to be the most suitable: \textbf{MA}, \textbf{UNH}, \textbf{INTC}, \textbf{NVDA}, and \textbf{HD}. The table below shows the head of the downloaded data frame for these assets.

\begin{table}[htbp]
\centering
\caption{Head of Adjusted Close Prices for Efficient Frontier Assets}
\label{tab:efficient-assets}
\scalebox{0.90}{%
\begin{tabular}{lrrrrr}
\toprule
Date & HD & INTC & MA & NVDA & UNH \\
\midrule
2010-01-04 & 20.428894 & 13.605182 & 23.820492 & 4.240229 & 25.497299 \\
2010-01-05 & 20.578539 & 13.598674 & 23.750000 & 4.302146 & 25.456848 \\
2010-01-06 & 20.507273 & 13.553061 & 23.715666 & 4.329666 & 25.707539 \\
2010-01-07 & 20.749550 & 13.422742 & 23.560692 & 4.244816 & 26.694113 \\
2010-01-08 & 20.649788 & 13.572608 & 23.569046 & 4.253988 & 26.443430 \\
\bottomrule
\end{tabular}%
}
\end{table}

\noindent All possible combinations of five assets were iterated by the model in order to identify the optimal asset mix through comparisons of expected returns, risk, and Sharpe ratios. The covariance matrix of the five optimal assets, from which the portfolio allocation was derived, was presented below.

\begin{table}[htbp]
\centering
\caption{Covariance Matrix for Optimal Assets}
\label{tab:cov-optimal}
\scalebox{0.90}{%
\begin{tabular}{lrrrrr}
\toprule
Ticker & MA & UNH & INTC & NVDA & HD \\
\midrule
MA   & 0.076418 & 0.034337 & 0.046560 & 0.068395 & 0.037899 \\
UNH  & 0.034337 & 0.067703 & 0.030882 & 0.039747 & 0.030644 \\
INTC & 0.046560 & 0.030882 & 0.111970 & 0.082922 & 0.039991 \\
NVDA & 0.068395 & 0.039747 & 0.082922 & 0.231897 & 0.054154 \\
HD   & 0.037899 & 0.030644 & 0.039991 & 0.054154 & 0.061295 \\
\bottomrule
\end{tabular}%
}
\end{table}

\noindent Before training the model, the discrete allocation was obtained as follows:

\begin{table}[htbp]
\centering
\caption{Discrete Allocation of Shares Prior to Model Training}
\label{tab:discrete-allocation}
\begin{tabular}{lr}
\toprule
\textbf{Asset} & \textbf{Number of Shares} \\
\midrule
INTC & 12 \\
MA   & 7  \\
UNH  & 4  \\
\bottomrule
\end{tabular}
\end{table}

\noindent Here, Intel (INTC) had the largest number of shares (12), followed by Mastercard (MA) with 7 shares, and UnitedHealth (UNH) with 4 shares. This allocation is illustrated in the Table~\ref{tab:discrete-allocation}.

\noindent The performance metrics of the portfolio before and after training the model are summarised below:

\begin{table}[htbp]
\centering
\caption{Portfolio Performance Metrics Before and After Model Training}
\label{tab:portfolio-performance}
\begin{tabular}{lcc}
\toprule
\textbf{Metric} & \textbf{Before Training} & \textbf{After Training} \\
\midrule
Expected/Annualized Return & 51.1\% & 2.10\% \\
Annual/Annualized Volatility & 34.9\% & 16.32\% \\
Sharpe Ratio & 1.41 & 0.13 \\
Winning Days Ratio & -- & 49.71\% \\
Information Ratio & -- & 3.96 \\
\bottomrule
\end{tabular}
\end{table}

\begin{figure}[H]
\centering
\includegraphics[width=0.5\textwidth]{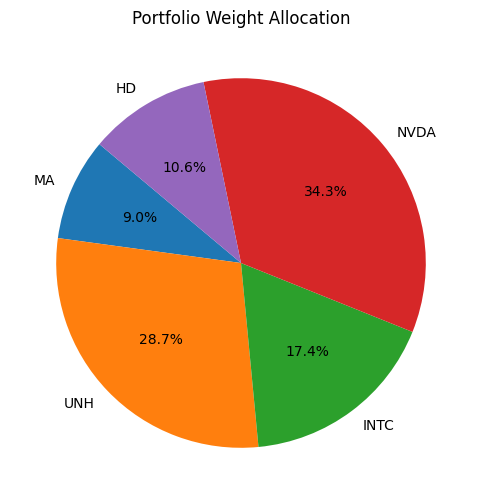}
\caption{The figure above shows the weight allocation for the equal weights portfolio and the DRL portfolio, which rebalances the weights at each time step or when it iterates using a policy that maximizes the portfolio’s cumulative value.}
\label{fig:portfolio-weights}
\end{figure}

\section{Results and Discussion}

\subsection{Performance Metrics}

The performance of the Deep Reinforcement Learning (DRL) portfolio allocation model was evaluated against its pre-training baseline. 
Table~\ref{tab:metrics} presents the three key metrics: annualized return, volatility, and Sharpe ratio.

\begin{table}[H]
\centering
\caption{Performance Metrics Before and After Training}
\label{tab:metrics}
\begin{tabular}{lcc}
\toprule
\textbf{Metric} & \textbf{Before Training} & \textbf{After Training} \\
\midrule
Annualized Return     & 51.1\% & 2.10\% \\
Annualized Volatility & 34.9\% & 16.32\% \\
Sharpe Ratio          & 1.41   & 0.13   \\
\bottomrule
\end{tabular}
\end{table}

\subsection{Result Analysis}

The baseline (pre-training) strategy delivered an annualized return of \textbf{51.1\%} with a Sharpe ratio of \textbf{1.41}, albeit at relatively high volatility (\textbf{34.9\%}). 
This indicates strong raw profitability but with significant exposure to risk. 
In contrast, after training, the DRL agent’s return declined sharply to \textbf{2.10\%}, with volatility reduced to \textbf{16.32\%}. 
The Sharpe ratio deteriorated to \textbf{0.13}, suggesting a collapse in risk-adjusted performance.

This shift highlights a critical trade-off: while the trained model successfully \textit{stabilized} risk through volatility reduction, it failed to preserve profitability. 
The decline in Sharpe ratio indicates that the model was unable to balance return generation against risk, and may have overfit to training data or converged to overly conservative portfolio allocations.

\subsection{Discussion}

Several factors may explain the observed performance degradation:
\begin{enumerate}
    \item \textbf{Overfitting to training data:} Reinforcement learning models often capture patterns specific to the training set, leading to poor generalization in out-of-sample backtests.
    \item \textbf{Exploration–exploitation imbalance:} The agent may have converged prematurely to low-risk allocations, sacrificing long-term return potential.
    \item \textbf{Non-stationarity of financial markets:} Changing market regimes can render learned policies suboptimal, especially if the model does not adapt to structural shifts.
\end{enumerate}

Despite these shortcomings, the post-training reduction in volatility is noteworthy: the agent implicitly learned risk control, albeit at the expense of returns. 
From a risk-aware optimization perspective, this demonstrates the model’s potential to align with investor preferences under certain utility frameworks, such as those prioritizing capital preservation over aggressive growth.

\subsection{Implications and Future Work}

The results emphasize the difficulty of applying DRL to real-world portfolio optimization. 
While volatility control was achieved, the collapse in Sharpe ratio reveals the necessity of:
\begin{itemize}
    \item \textbf{Reward shaping:} Incorporating risk-adjusted objectives, such as maximizing the Sharpe ratio or Conditional Value-at-Risk (CVaR), to better align learning signals with portfolio goals.
    \item \textbf{Robust validation:} Employing walk-forward testing and cross-validation across multiple market regimes to mitigate overfitting.
    \item \textbf{Risk-aware constraints:} Integrating drawdown limits, turnover penalties, and diversification constraints to prevent convergence to trivial low-risk strategies.
\end{itemize}

Overall, while the trained agent underperformed relative to the baseline in absolute and risk-adjusted terms, the findings underscore the importance of explicitly encoding \textit{risk-aware optimization} into the reinforcement learning framework. 
Future research should focus on hybrid models that combine DRL with financial domain knowledge, adaptive learning mechanisms, and interpretable reward structures.

\section{Conclusion}

This study explored the use of deep reinforcement learning (DRL) with risk-aware optimization for dynamic portfolio allocation in finance. The empirical results demonstrate both the potential and the limitations of such approaches. Prior to training, the DRL model exhibited a strong performance with an annualized return of 51.1\%, volatility of 34.9\%, and a Sharpe ratio of 1.41, indicating favorable risk-adjusted performance. However, after training, performance deteriorated significantly, with the Sharpe ratio declining to 0.13, annualized return falling to 2.1\%, and volatility reducing to 16.32\%. 

The post-training results suggest that the model may have suffered from overfitting, instability in policy convergence, or inadequate alignment between the reward function and long-term risk-return trade-offs. Despite this, the model achieved a winning days ratio of 49.71\% and an information ratio of 3.96, showing that while daily consistency improved modestly, the cumulative performance failed to sustain pre-training levels.

Overall, the findings highlight that while DRL has the capacity to generate competitive strategies in portfolio optimization, careful design of reward structures, regularization against overfitting, and enhanced exploration-exploitation balance are essential for stability and long-term viability. Future research should focus on integrating advanced risk measures, hybrid models that combine DRL with econometric predictors, and extensive cross-validation across different market regimes. This would help to bridge the gap between promising pre-training simulations and robust, real-world deployment of DRL-based portfolio strategies.

\bibliographystyle{unsrtnat}
\bibliography{references}  






\end{document}